\documentclass[12pt]{article}
\usepackage{amsmath,wasysym}
\usepackage{amssymb}
\usepackage{graphicx}
\usepackage{float}
\usepackage{indentfirst}
\usepackage{mathrsfs}
\usepackage{dsfont}

\usepackage{setspace}

\usepackage[labelfont=bf,labelsep=period]{caption}

\usepackage[numbers,sort&compress]{natbib}

\newtheorem{proposition}{Proposition}
\newtheorem{corollary}{Corollary}

\title{An Optimal Measurement Strategy to Beat the Quantum Uncertainty in Correlated System\\ [0.7cm]}

\author
{Jun-Li Li and Cong-Feng Qiao$^{\ast}$ \\ [0.2cm]
\normalsize{School of Physical Sciences, University of Chinese Academy of Sciences,} \\
\normalsize{YuQuan Road 19A, Beijing 100049, China}\\ [2pt]
\normalsize{Center of Materials Science and Optoelectronics Engineering \& CMSOT,} \\
\normalsize{University of Chinese Academy of Sciences, YuQuan Road 19A, Beijing 100049, China}\\ [2pt]
\normalsize{Key Laboratory of Vacuum Physics, University of Chinese Academy of Sciences} \\
\normalsize{YuQuan Road 19A, Beijing 100049, China} \\ [3mm]
\normalsize{$^\ast$ To whom correspondence should be addressed; E-mail: qiaocf@ucas.ac.cn.}
}

\date{}

\begin{document}
\baselineskip24pt \maketitle
\begin{abstract}\doublespacing
Uncertainty principle is an inherent nature of quantum system that undermines the precise measurement of incompatible observables and hence the applications of quantum theory. Entanglement, another unique feature of quantum physics, was found may help to reduce the quantum uncertainty. In this paper, we propose a practical method to reduce the one party measurement uncertainty by determining the measurement on the other party of an entangled bipartite system. In light of this method, a family of conditional majorization uncertainty relations in the presence of quantum memory is constructed, which is applicable to arbitrary number of observables. The new family of uncertainty relations implies sophisticated structures of quantum uncertainty and nonlocality, that were usually studied by using scalar measures. Applications to reduce the local uncertainty and to witness quantum nonlocalities are also presented.
\end{abstract}

\newpage

\section{Introduction}

One of the distinct features of quantum mechanics is its inherent limit on the joint measurement precisions of incompatible observables, known as the uncertainty relation. The most representative uncertainty relation writes \cite{Robertson}:
\begin{equation}
\Delta X \Delta Y \geq \frac{1}{2}|\langle [X,Y] \rangle| \; . \label{Robertson-Uncertainty}
\end{equation}
Here the product of the uncertainties of two observables $X$ and $Y$ is lower bounded by the expectation value of their commutator. This lower bound is state dependent and could be null which trivializes the uncertainty relation \cite{Entropy-UN-MU-1}. The entropic form of uncertainty relations may avoid such problem, which were developed with the state independent lower bounds \cite{Entropy-UN-MU-1, Entropy-UN-MU-2, Entropy-UN-MU-3}. A typical one of them, the Maassen and Uffink(MU) form \cite{Entropy-UN-MU-3}, goes as
\begin{equation}
H(X)+H(Y) \geq \log \frac{1}{c} \; . \label{Entropic-MU-Relation}
\end{equation}
Here $H(X)=-\sum_i p_i\log p_i$ denotes the Shannon entropy, $p_i$ represents the probability of obtaining $x_i$ while measuring $X$, and similarly for observable $Y$. The logarithm is assumed to have base number $e$ if not further specified, and the symbol $c \equiv \max_{i,j} |\langle x_i|y_j\rangle|^2$ quantifies the maximal overlap of eigenvectors $|x_i\rangle$ and $|y_j\rangle$ of $X$ and $Y$, respectively. Though great efforts have been made, to find out the optimal lower bound for entropic type uncertainty relation remains a challenging task \cite{Entropy-RMD}. It is remarkable that according to a recent study the variance and entropy types of uncertainty relations in fact are mutually equivalent \cite{Equiv-V-E}.

Recently, people found that though uncertainty is an inherent nature of quantum physics, it is beatable in the presence of quantum memory \cite{Uncertainty-Memory-1}. In such situation, the uncertainty relation takes the following form
\begin{equation}
S(X|B)+S(Y|B) \geq \log \frac{1}{c} + S(A|B) \; . \label{entropy-memory-equ1}
\end{equation}
Here $S(X|B)$ stands for conditional von Neumann entropy, representing the uncertainty in the measurement of $X$ on Alice($A$) side, given information is stored as quantum memory on Bob($B$). The term $S(A|B)$ on the right hand side of equation (\ref{entropy-memory-equ1}) is supposed to signify the influence of entanglement on the uncertainty relation, but actually it has no business with the local measurement uncertainties, viz. $H(X)$ or $H(Y)$. In the literature, though a lot of effort have been made \cite{Entropy-rev-ADP, Entropy-tri-mem}, equation (\ref{entropy-memory-equ1}) is still subject to the absence of optimal lower bound in entropic uncertainty relation \cite{Uncertainty-Memory-2}.

Contrary to the variance and entropy, a vectorized measure of uncertainty was introduced in \cite{Major-intro-1}, where the majorization employed helps to improve the entropic uncertainty relation \cite{Major-intro-2} and leads to a universal uncertainty relation \cite{Major-intro-3}. A typical majorization uncertainty relation in direct sum form goes as \cite{Maj-Un-Relation}
\begin{align}
\vec{p}(x) \oplus \vec{p}(y)  \prec \vec{s} \;, \label{Conditional-maj-1}
\end{align}
where $\vec{p}(\cdot)$ signifies the probability distribution of the  measurement outcomes. The majorization relation $\vec{a} \prec \vec{b}$ is defined as $\sum_{i=1}^ka_i\leq \sum_{j=1}^kb_j$, $k\in\{1,\cdots, N\}$ for two vectors with components in descending order and the equality holds for $k=N$. Unlike variance and entropy, the upper bound $\vec{s}$ in uncertainty relation (\ref{Conditional-maj-1}) is unique and optimal which can be easily determined via the theory of majorization lattice. Whereas, it was proved that this limit may be violated in the situation with entangled bipartite state \cite{Quantum-Magic}.

For a bipartite system, it is interesting to know  what measurement strategy $B$ takes can minimize the measurement uncertainty on $A$'s side. Such an optimal measurement strategy would build a maximal correlation between measurements, from which a substantial reduction on local uncertainty can be deduced. In addition to reducing the local uncertainty, one can imagine that the correlation maximality would have certain effects on other quantum information processing scenarios. For example, the Bell nonlocality, quantum steering \cite{Steer-1}, and non-separability may be somehow distinguished by stronger correlations reached by optimal joint measurements in bipartite system. In statistical inference, a typical problem is to identify the extremal joint distribution that maximizes the correlation for given marginals \cite{Mini-Ent-App}, which is closely related to infer a maximal correlation for given joint measurements.

In this paper, we propose a practical method to reduce the local measurement uncertainty in the presence of quantum memory. That is, given the bipartite quantum state $\rho_{AB}$ and measurement $X$ on $A$, how to perform the measurement $X'$ on $B$ to reduce the uncertainty in measuring $X$ on $A$. By virtue of the lattice theory, an optimal measurement strategy will be established. To this aim, we formulate a family of conditional majorization uncertainty relations (CMURs) in the presence of quantum memory. The CMUR is applicable to multiple, including infinitely many, observables, which enables us to study the uncertainty relation and nonlocality with infinite number of measurement settings. This is a hurdle hard to surmount in other formalisms, i.e., variance or entropic uncertainty relations are difficult to adapt to infinitely many observables.

\section{Optimal measurement strategy and applications}

\subsection{Optimal measurements to reduce the local uncertainty}

In a bipartite system $\rho_{AB}$, the reduced density matrix $\rho_{A} = \mathrm{Tr}_B[\rho_{AB}]$ describes the local system $A$. The diagonal elements of $\rho_{A}' = u_x^{\dag}\rho_A u_x$ give the outcome distribution of the measurement $X$, where the unitary matrix $u_x= (|x_1\rangle, \cdots, |x_N\rangle)$ is composed of the eigenvectors of $X$. A measurement on system $B$, i.e., $X'$ of dimension $N$, may also be performed, and hence results in a joint distribution
\begin{align}
P(X,X') = ( \vec{p}^{\,(1)}(x) , \vec{p}^{\,(2)}(x) ,\cdots, \vec{p}^{\,(N)}(x)) \; .
\end{align}
Here $\vec{p}^{\,(\mu)}(x)$ represents the distribution vector of the measurement $X$ conditioned on $X'$ being found to be $x_{\mu}'$. For the joint distribution $P(X,X')$, the $X$'s marginal distribution is $\vec{p}(x)=\sum_{\mu =1}^N \vec{p}^{\,(\mu)}(x)$. We define the vectorized measure of uncertainty for $X$ conditioned on the knowledge of $X'$ as
\begin{align}
\vec{p}(x|x') \equiv \vec{p}^{\,(1)\downarrow} (x) + \vec{p}^{\,(2)\downarrow} (x) + \cdots +\vec{p}^{\,(N)\downarrow} (x)\; , \label{Def-cond-XX}
\end{align}
where the superscript $\downarrow$ indicates that the components of the summand vectors are arranged in descending order. The conditional probability distribution $\vec{p}(x|x')$ may be called the {\it majorized marginal distribution}. For two joint distributions $P(X,X')$ and $P(X,Y')$, $\vec{p}(x|x') \prec \vec{p}(x|y')$ means that $X$ has less uncertainty conditioned on the information of $X'$ than that conditioned on $Y'$.

With the definition of majorized marginal distribution, one may figure out that, 1. $\vec{p}(x|x')$ is always less uncertain than the ordinary marginal $\vec{p}(x)$, i.e., $\forall X'$, $\vec{p}(x) \prec \vec{p}(x|x')$; 2. The joint distribution of independent observables $X$ and $X'$ gives $\vec{p}(x) = \vec{p}(x|x')$. For all possible measurements $X'$, the majorized marginal distributions $\vec{p}(x|x')$ have the following property:
\begin{proposition}
Given $\rho_{AB}$ and the measurement $X$ on $A$, there exists a unique least upper bound for $\vec{p}(x|x')$, i.e.,
\begin{align}
\forall X'\;, \; \vec{p}(x|x') \prec \vec{s}^{\,(x)} \; . \label{Maj-optimal-Measure}
\end{align}
Here $\vec{s}^{\,(x)}= \vec{p}(x|x'^{(1)})\vee \vec{p}(x|x'^{(2)}) \vee \cdots \vee \vec{p}(x|x'^{(N)})$ depends only on $X$ and $\rho_{AB}$. The majorized marginal $\vec{p}(x|x'^{(k)})$ denotes the distribution vector having the largest sum of the first $k$ components, i.e. $\sum_{i=1}^k p_i(x|x'^{(k)}) = \max\limits_{\{X'\}} \left\{\sum_{i=1}^k p_i(x|x')\right\}$. \label{Proposition-1}
\end{proposition}
Majorization relation (\ref{Maj-optimal-Measure}) holds due to the fact that there exists a least upper bound for the join operator `$\vee$' of a majorization lattice \cite{Maj-Un-Relation}. The unique least upper bound $\vec{s}^{\,(x)}$ is determined by the measurement set $\left\{X'^{(k)}|k=1,\cdots,N\right\}$, in which each $X'^{(k)}$ gives $P(X,X'^{(k)})$ and then the marjorized marginal distribution of $\vec{p}(x|x'^{(k)})$ is defined from equation (\ref{Def-cond-XX}) (see Appendix A). We call the measurement set $\{X'^{(k)}\}$ the optimal measurement strategy for $B$ to reduce the uncertainty of $X$. In the following, we present several typical applications based on this measurement strategy.

\subsection{The conditional majorization uncertainty relation (CMUR)}

The uncertainty relations behave as the constraints on the probability distributions of two or more incompatible measurements. Variance and entropy are scalar measures of the distribution uncertainty (disorder or randomness in the language of statistics), while the majorization relation provides a lattice structured uncertainty measure \cite{Maj-Un-Relation}. For bipartite system, the measurement uncertainty of one party may be reduced based on its entangled partner's side information. Considering of equation (\ref{Maj-optimal-Measure}) we have the following Corollary
\begin{corollary}
For two measurements $X$ and $Y$ on $A$, we have a family of CMURs
\begin{align}
\vec{p}(x|x') * \vec{p}(y|y') \prec \vec{s}^{\,(*)} \; .  \label{Bi-maj-conditional}
\end{align}
Here $X'$ and $Y'$ are measurements on $B$, and $\vec{s}^{\,(*)} \equiv \vec{s}^{\,(x)} * \vec{s}^{\,(y)}$ with $\{*\}$ being a set of binary operations that preserve the majorization which include direct sum, direct product, and vector sum, i.e., $\{\oplus,\otimes, +\} \subset \{*\}$. \label{Corollary-1}
\end{corollary}
The proof of Corollary \ref{Corollary-1} is deferred to the Appendix B for simplicity. In fact, the relation (\ref{Bi-maj-conditional}) is not restricted to observable pairs. Arbitrary number of observables apply here by the multiple application of the operations in $\{*\}$. An alternative definition of conditional majorization can be found in \cite{Cond-maj-18}, where convex functions are used explicitly for the characterization. Next, we show how the CMUR presented here affects quantum measurements, namely on the uncertainty relation with entangled quanta and the witness of quantum steering.

\subsubsection{Break the quantum constraint}

\begin{figure}[t]\centering
\scalebox{0.37}{\includegraphics{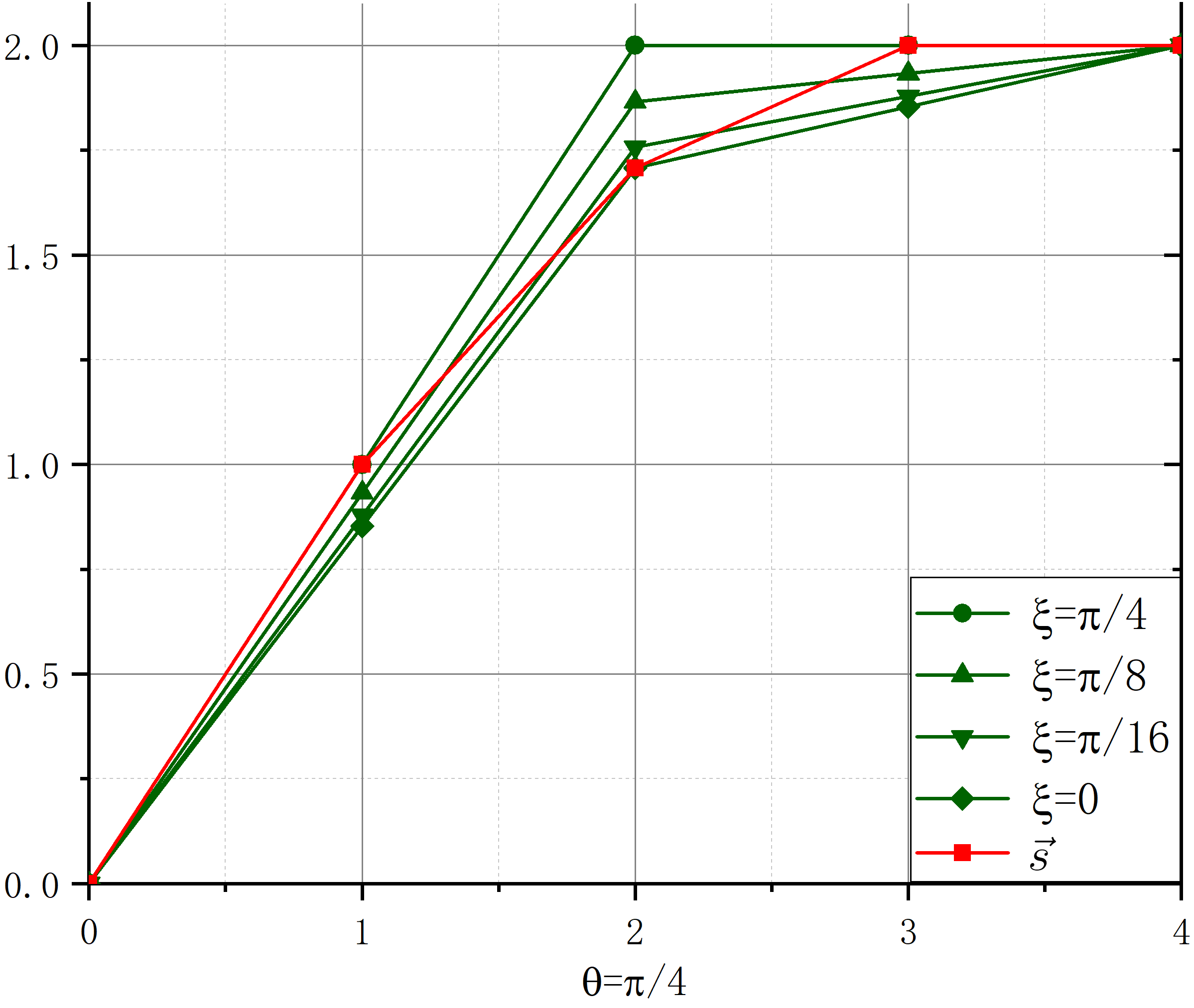}}\;\;
\scalebox{0.37}{\includegraphics{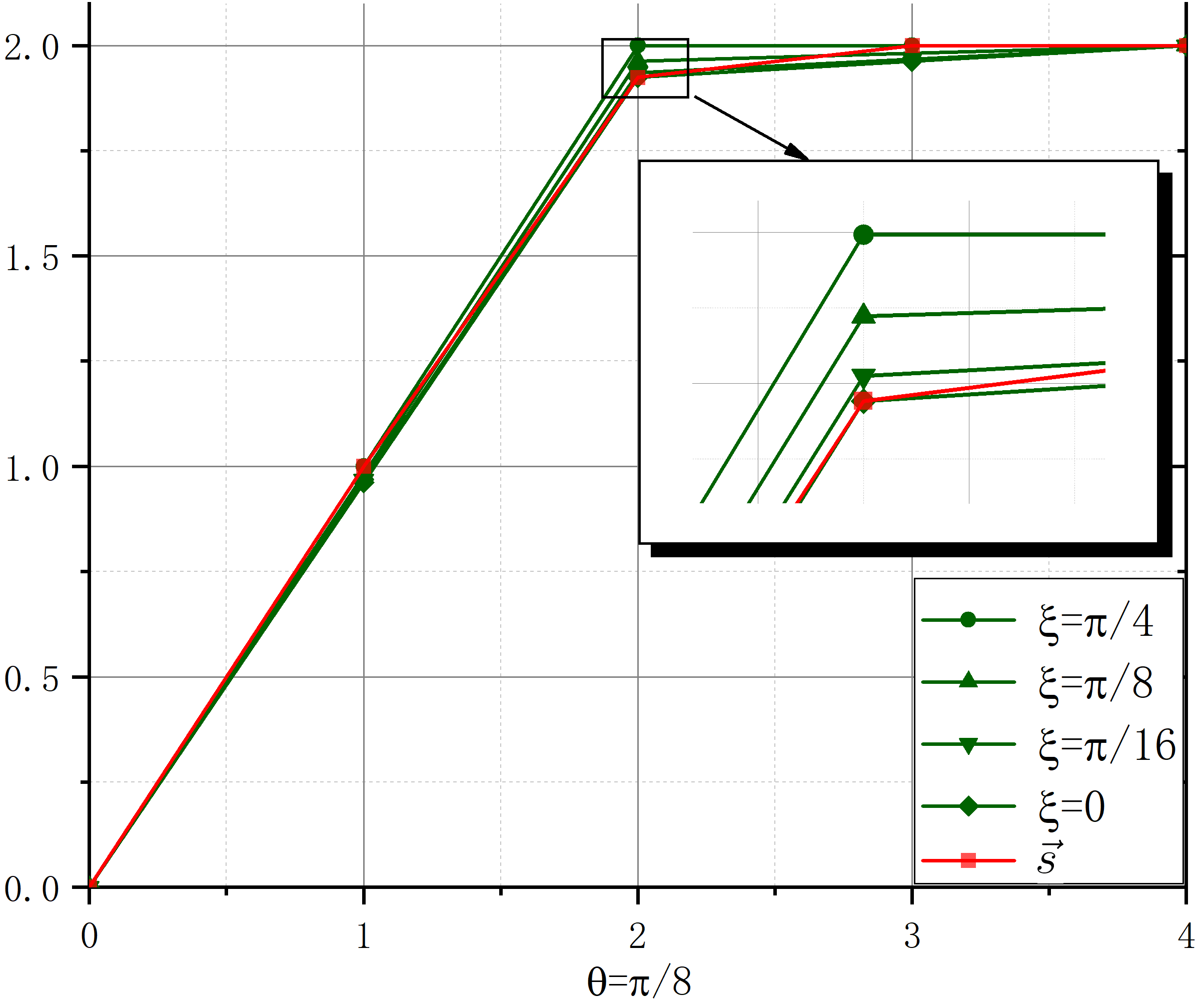}}
\caption{{\bf The conditional majorization uncertainty relation in the presence of quantum memory.} The violation of the quantum bound $\vec{s}$ for observable pairs $\{\sigma(\theta,0), \sigma(\theta,\pi)\}$ with different $\theta$ are presented for the quantum state $|\psi_{\xi}\rangle = \cos \xi |00\rangle + \sin\xi |11\rangle $. The green lines represent the Lorenz curves of the upper bound $\vec{s}^{\,(\oplus)}$ for different $\xi$, which violate the quantum limit $\vec{s}$, the red lines, in certain sections.} \label{Figure-UR-Maj}
\end{figure}

Considering the operation of direct sum in Corollary \ref{Corollary-1}, we have
\begin{align}
\vec{p}(x|x') \oplus \vec{p}(y|y') \prec \vec{s}^{\,(\oplus)} = \vec{s}^{\,(x)}\oplus \vec{s}^{\,(y)}\; . \label{Direc-sum-XY-bounds}
\end{align}
This provides an upper bound for incompatible measurements $X$ and $Y$ on $A$ in the presence of quantum memory. We know for single particle the optimal upper bound for the direct sum majorization uncertainty relation of $X$ and $Y$ is \cite{Maj-Un-Relation}
\begin{align}
\vec{p}(x) \oplus \vec{p}(y) \prec \vec{s} \; ,
\end{align}
where $\vec{s}$ depends only on the local observables. To compare these two bounds of $\vec{s}^{\,(\oplus)}$ and $\vec{s}$, we employ the bipartite qubit state
\begin{align}
|\psi_{\xi}\rangle = \cos \xi |00\rangle + \sin\xi |11\rangle \; , \; \xi\in [0,\pi/4]\;  \label{psi-xi}
\end{align}
and local observable
\begin{align}
\sigma_n \equiv \sigma(\theta,\phi) = \sigma_z\cos\theta + \sigma_x\sin\theta\cos\phi + \sigma_y\sin\theta\sin\phi \;.
\end{align}
According to Proposition \ref{Proposition-1}, for measurement $\sigma_n$ applying on $A$, the optimal measurement $\sigma_{n'}=\sigma(\theta',\phi')$ performed by $B$ is achieved by taking (see Appendix A for details)
\begin{align}
\tan\theta' = \tan\theta\sin(2\xi)\;,\; \phi'=\phi \; ,
\end{align}
and the corresponding optimal upper limit for the majorized marginal distribution is
\begin{align}
\vec{p}(\sigma_n|\sigma_{n'}) \prec \vec{s}^{\,(\sigma_n)} = \begin{pmatrix}
\frac{1}{2}+\frac{1}{2}\sqrt{\cos^2\theta+\sin^2\theta\sin(2\xi)} \\
\frac{1}{2}-\frac{1}{2}\sqrt{\cos^2\theta+\sin^2\theta\sin(2\xi)}
\end{pmatrix} \;. \label{Pxx-optimal}
\end{align}
Taking $X=\sigma(\theta,0)$ and $Y=\sigma(\theta,\pi)$ with commutator $[X,Y] = i\sigma_y\sin(2\theta)$, the upper bound $\vec{s}^{\,(\oplus)}$ in relation (\ref{Direc-sum-XY-bounds}) can be constructed via (\ref{Pxx-optimal}). The relationship between $\vec{s}^{\,(\oplus)}$ (with the presence of quantum memory) and $\vec{s}$ (single particle state) for different degrees of entanglement (characterized by $\xi$) is illustrated in Figure \ref{Figure-UR-Maj} in Lorenz curves. Two distributions satisfy $\vec{s}^{\,(\oplus)}\prec \vec{s}$, if and only if the Lorenz curve of $\vec{s}^{\,(\oplus)}$ is everywhere below that of $\vec{s}$. The state $|\psi_{\xi}\rangle$ is entangled when $\xi>0$ and reaches the maximal entanglement at $\xi=\pi/4$. In the whole range of $\xi \in (0,\pi/4]$, the CMUR (\ref{Direc-sum-XY-bounds}) has the upper bound $\vec{s}^{\,(\oplus)}\nprec \vec{s}$, see Figure \ref{Figure-UR-Maj}. That is, the quantum limit can be broken in the presence of entanglement.

\subsubsection{Compare to the conditional entropic uncertainty relation}

\begin{figure}\centering
\scalebox{0.6}{\includegraphics{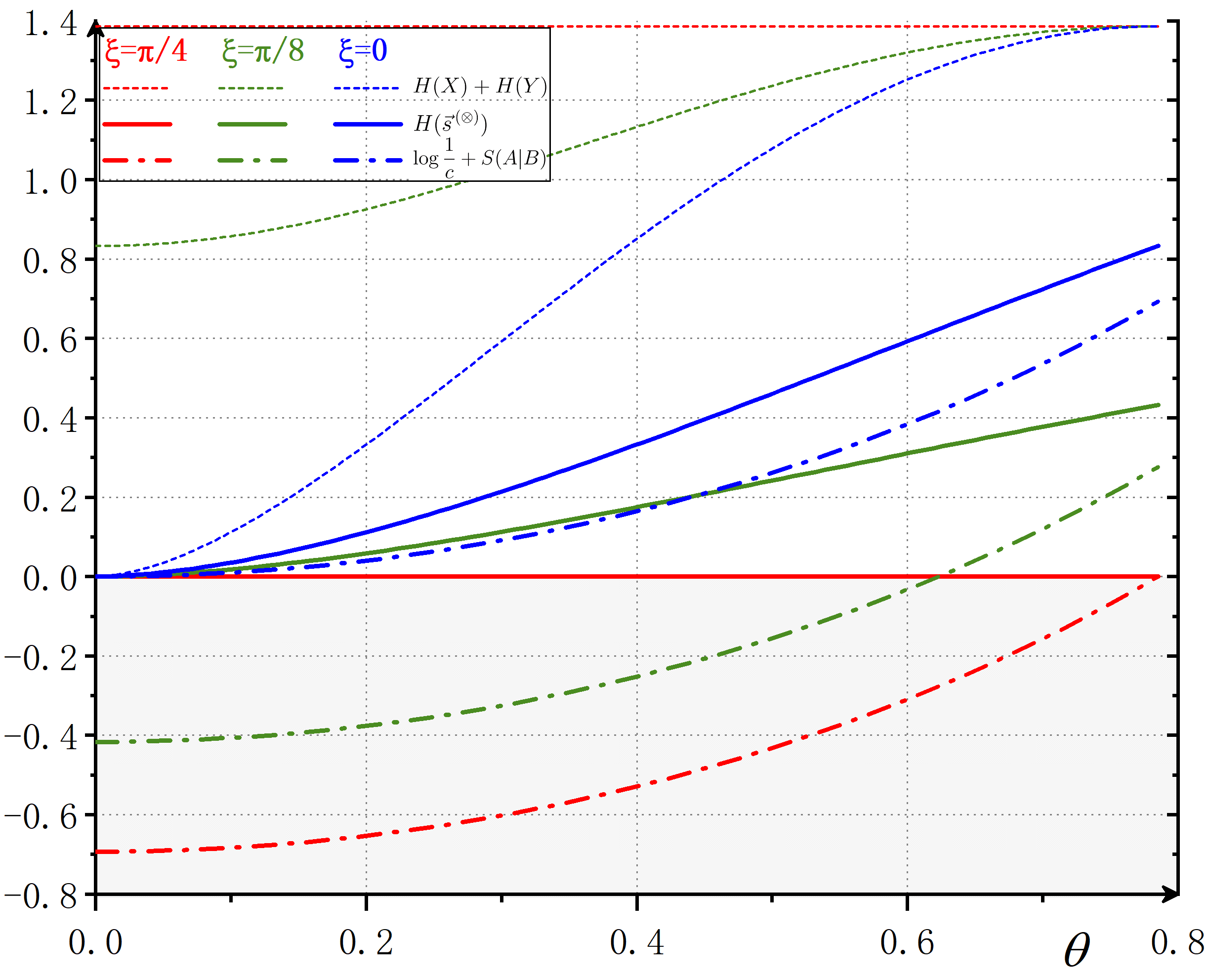}}
\caption{{\bf The decrease of the lower bound for the entropic uncertainty relation with quantum memory.} The uppermost dashed lines give the values of $H(X)+H(Y)$ for $\rho_A = \mathrm{Tr}_B[|\psi_{\xi}\rangle \langle \psi_{\xi}|]$. The solid lines represent the reduced lower bounds of $H(\vec{s}^{\,(\otimes)})$ under the optimal measurement strategy, and the dot-dashed lines represent the lower bounds of $\log\frac{1}{c}+S(A|B)$. The lower bounds are plotted for $\xi=\frac{\pi}{4}, \frac{\pi}{8}, 0$ in different colors which correspond to different degrees of entanglement of $|\psi_{\xi}\rangle$. The lower bounds $H(\vec{s}^{\,\otimes})$ are greater (thus tighter) than $\log\frac{1}{c}+S(A|B)$ for each value of $\xi$, respectively. Note, a large portion of the dot-dashed lines are negative and hence meaningless. } \label{Figure-conditional-entropy}
\end{figure}

To compare with the existing result on the entropic uncertainty relation in the presence of quantum memory, the relation (\ref{entropy-memory-equ1}), we transform the CMUR (\ref{Bi-maj-conditional}) into the entropic form. In practice this is quite straightforward by applying an arbitrary Schur's concave function to $\vec{p}(x|x')\prec \vec{s}^{\,(x)}$ and $\vec{p}(y|y')\prec \vec{s}^{\,(y)}$. For the Shannon entropy $H(\cdot)$ on the direct product form of equation (\ref{Bi-maj-conditional}), we have
\begin{align}
H\left(\vec{p}(x|x') \right) + H\left(\vec{p}(y|y')\right) & \geq H\left(\vec{s}^{\,(\otimes)} \right) \; . \label{Reduced-Shan-entropy}
\end{align}
Here $\vec{s}^{\,(\otimes)} = \vec{s}^{\,(x)} \otimes \vec{s}^{\,(y)}$.
Figure \ref{Figure-conditional-entropy} exhibits the behaviors of the lower bounds in equations (\ref{entropy-memory-equ1}) and (\ref{Reduced-Shan-entropy}) with the quantum state $|\psi_{\xi}\rangle$ being the form of (\ref{psi-xi}) and observables $X=\sigma(\theta,0)$ and $Y=\sigma(\theta,\pi)$. The uncertainties of $X$ and $Y$ on $A$'s side are evaluated for the reduced density matrix $\rho_A = \mathrm{Tr}_B [|\psi_{\xi}\rangle \langle \psi_{\xi}]$. Note, $H(X)+H(Y)$ for $\rho_A$ is always larger than zero. However, when beating the uncertainties with quantum memory of $|\psi_{\xi}\rangle$ based on equation (\ref{entropy-memory-equ1}), the right hand side of equation (\ref{entropy-memory-equ1}) is $\log \frac{1}{c} + \mathrm{Tr}[\rho_B\log\rho_B]$ with $\rho_B = \mathrm{Tr}_A[|\psi_{\xi}\rangle \langle \psi_{\xi}|]$, and the lower bounds are mostly negative (dot-dashed lines in Figure \ref{Figure-conditional-entropy}) for the parameters $\xi\in [0,\pi/4]$ in this case. In comparison, the reduced lower bounds for local uncertainties in equation (\ref{Reduced-Shan-entropy}) are realistic and physically reachable (the solid lines in Figure \ref{Figure-conditional-entropy}).

\subsection{Uncertainty relaiton with infinite number of observables}

The Corollary \ref{Corollary-1} can be easily generalized to arbitrary number of observables. Taking the vector sum operation as an example, for $M$ observables $X^{(i)}$ on $A$ we have
\begin{align}
\sum_{i=1}^M \vec{p}(x^{(i)}|x'^{(i)}) \prec \vec{s}^{\,(+)}\; , \label{Sum-maj-M}
\end{align}
where $\vec{s}^{\,(+)} = \sum_{i=1}^M \vec{s}^{\,(x_i)\downarrow}$. Relation (\ref{Sum-maj-M}) represents a CMUR in the presence of quantum memory for $M$ incompatible observables. For single particle states, there is \cite{Maj-Un-Relation}
\begin{align}
\bigoplus_{i=1}^M \vec{p}(x^{(i)}) \prec \vec{s} \; . \label{Maj-Directsum-NM}
\end{align}
Here $\vec{s}$ is a real vector of dimension $NM$ with components arranged in descending order. By partitioning the vector $\vec{s}$ into $N$ disjoint sections as $\vec{s} = (s_1,\cdots, s_M; s_{M+1},\cdots, s_{2M}; \cdots)$, we may get an $N$-dimensional vector $\vec{\varepsilon}$ with the components of $\varepsilon_i = \sum_{j=1}^M s_{ (i-1)\cdot M+j}$. Then relation (\ref{Maj-Directsum-NM}) leads to
\begin{align}
\sum_{i=1}^M \vec{p}(x^{(i)}) \prec \vec{\varepsilon} \; , \label{s-aggregation}
\end{align}
where the vector $\vec{\varepsilon}$ is called the aggregation of $\vec{s}$ \cite{Mini-Ent-App}. Relation (\ref{s-aggregation}) represents the quantum prediction of relation (\ref{Sum-maj-M}) without quantum memory.

The violation of the relation $\vec{s}^{\,(+)} \prec \vec{\varepsilon}$ implies a violation of relation (\ref{Maj-Directsum-NM}), which then implies the quantum steering in a bipartite state \cite{Quantum-Magic}. Defining the correlation matrix of a bipartite state $\rho_{AB}$ as as $\mathcal{T}_{ij} = \mathrm{Tr}[\rho_{AB} \sigma_{i} \otimes \sigma_{j}]$ where $\sigma_{i}$ are three Pauli operators, the CMUR (\ref{Sum-maj-M}) leads to the following Corollary
\begin{corollary}
If a bipartite qubit state $\rho_{AB}$ is non-steerable, then
\begin{align}
R_{G}(\tau_3^2,\tau_2^2,\tau_1^2\,) \leq \frac{1}{2} \; .
\end{align}
Here $R_{G}$ signifies certain Elliptic integral and $\tau_1\geq \tau_2\geq \tau_3$ are singular values of the correlation matrix $\mathcal{T}$ of $\rho_{AB}$. \label{Corollary-2}
\end{corollary}
The definition of $R_{G}$ can be found in Ref. \cite{Special-Fun-Ell} (No.19.16.3 DLMF of NIST and is also presented in Appendix C).

We consider a typical mixed and entangled state \cite{Steering-Example-1}
\begin{align}
\rho_{\xi} & = \frac{1-p}{2}\rho^{A}_{\xi} \otimes \mathds{1} + p|\psi_{\xi}\rangle \langle \psi_{\xi}| \; , \; p\in[0,1] \; ,
\end{align}
where $\rho^A_{\xi} = \mathrm{Tr}_B[|\psi_{\xi}\rangle\langle \psi_{\xi}|]$ and $B$ can steer $A$ whenever $p>1/2$. For measurement $X=\sigma(\theta, \phi)$ on $A$, the Proposition \ref{Proposition-1} tells that the optimal measurement $X'=\sigma(\theta', \phi')$ for $B$ to reduce the local uncertainty of $X$ is
\begin{align}
\left\{\begin{array}{l}
\mathrm{if}\; \cos\theta \in [0, \frac{p\tan(2\xi)}{(1-p^2)^{1/2}}],\; \mathrm{then} \; \tan\theta' = \tan\theta\sin(2\xi)\; \mathrm{and}\; \phi'=\phi \\ \\  \mathrm{if}\; \cos\theta \in [\frac{p\tan(2\xi)}{(1-p^2)^{1/2}},1] ,\; \mathrm{then} \; \theta', \phi'\; \text{can be arbitrary}
\end{array} \right.\;.
\end{align}
On the other hand, for measurement $X'=\sigma(\theta', \phi')$ on $B$, the optimal measurement $X=\sigma(\theta, \phi)$ of $A$ to reduce the uncertainty of $X'$ is (see Appendix D)
\begin{align}
\text{if}\; \cos\theta' \in [0,1],\; \mathrm{then} \; \tan\theta = \tan\theta'\sin(2\xi)\; \mathrm{and} \; \phi=\phi' \;.
\end{align}
Evidently, the bipartite state $\rho_{\xi}$ is asymmetric from the optimal measurement point of view, though Corollary \ref{Corollary-2} is insensitive to this asymmetry.

\begin{figure}\centering
\scalebox{0.5}{\includegraphics{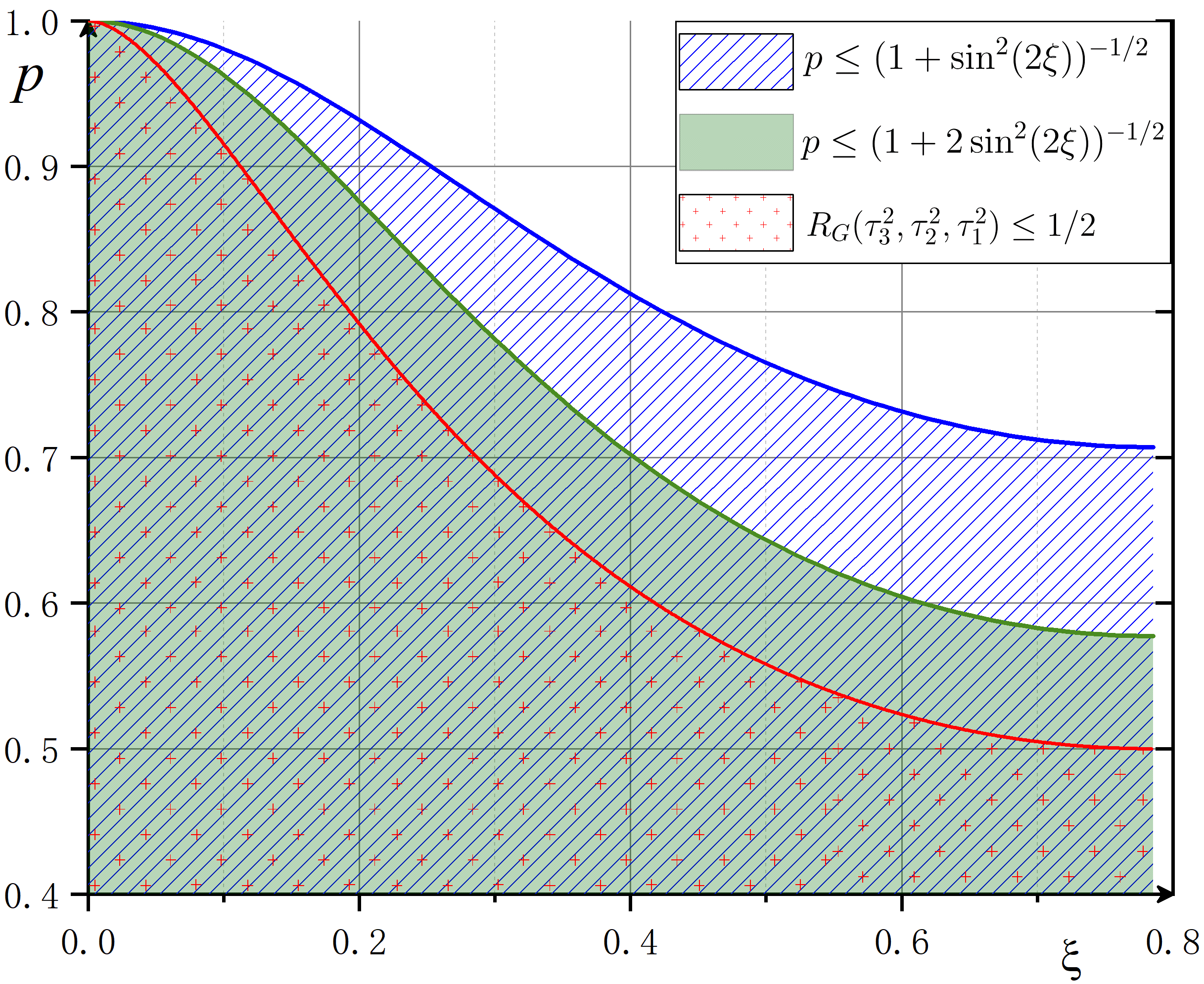}}
\caption{{\bf The criterion of steerability for two, three, and infinite number of projective measurements.} For state $\rho_{\xi}$, if $A$ cannot steer $B$, then we have: $p\leq (1+\sin^2(2\xi))^{-1/2}$ for two-measurement settings; $p\leq (1+2\sin^2(2\xi))^{-1/2}$ for three-measurement settings; $R_{G}(\tau_2^2, \tau_2^2, \tau_1^2)\leq 1/2$ for infinite-measurement settings.}
\label{Figure-3-criterion}
\end{figure}

It is remarkable that the steerability is greatly improved in the case of infinite number of measurements, which is illustrated in Figure \ref{Figure-3-criterion}. For state $\rho_{\xi}$, with infinite number of measurements Corollary \ref{Corollary-2} tells that if $A$ cannot steer $B$, the parameters $\xi$-$p$ will fall into the region of $R_{G}(\tau_3^2,\tau_2^2,\tau_1^2\,) \leq 1/2$ in Figure \ref{Figure-3-criterion}, where $\tau_i$ are the singular values of the correlation matrix of $\rho_{\xi}$. And in two- and three-measurement settings \cite{G-entropy-steering}, $A$ cannot steer $B$ predicts a much larger region than that of infinite-measurement setting. The mismatch of those areas indicates that although the state does not show steerability in less-measurement settings, it turns out to be steerable when more measurements are taken.

\section{Conclusion}

In conclusion, by virtue of the majorization lattice, we developed a systematic procedure to reduce the local uncertainties via entanglement. On account of this scheme, a practical measurement strategy was proposed and also a new class of conditional majorization uncertainty relations (CMUR) was constructed, which is applicable to any number of observables. In the presence quantum memory, it is found that the CMUR may break the quantum constraints on the measurement uncertainties of individual quanta. That is, a substantial reduction of local uncertainty on one partite can be fulfilled by performing some designated measurements on its entangled partner. Comparing to the conditional entropic uncertainty relation, the CMUR in its entropic forms may give out reduced and physically nontrivial lower bounds. Moreover, the CMUR can be employed to witness the steerability of bipartite states with infinite measurement settings, to which the usual uncertainty relations, like variance or entropy type of uncertainty relation, are inapplicable. Last, the measure of uncertainty we constructed is a novel formalism in materializing the uncertainty principle of quantum theory, we believe its application to quantum processing deserves a further investigation.

\section*{Acknowledgements}
\noindent
This work was supported in part by the Strategic Priority Research Program of the Chinese Academy of Sciences, Grant No.XDB23030100; and by the National Natural Science Foundation of China(NSFC) under the Grants 11975236 and 11635009; and by the University of Chinese Academy of Sciences.

\newpage
\setcounter{figure}{0}
\renewcommand{\thefigure}{S\arabic{figure}}
\setcounter{equation}{0}
\renewcommand\theequation{S\arabic{equation}}
\setcounter{theorem}{0}
\renewcommand{\thetheorem}{S\arabic{theorem}}
\setcounter{observation}{0}
\renewcommand{\theobservation}{S\arabic{observation}}
\setcounter{proposition}{0}
\renewcommand{\theproposition}{S\arabic{proposition}}
\setcounter{lemma}{0}
\renewcommand{\thelemma}{S\arabic{lemma}}
\setcounter{corollary}{0}
\renewcommand{\thecorollary}{S\arabic{corollary}}
\setcounter{section}{0}
\renewcommand{\thesection}{S\arabic{section}}

\appendix{\bf \LARGE Appendix}

We present detailed proofs of the Propositions and Corollaries in the text.

\section{Demonstration of Proposition 1}

For given bipartite system $\rho_{AB}$ and measurement $X$ of $A$, we may get the assemblages for $B$ \cite{S-Assem-d}
\begin{align}
\rho_{i|x} = \langle x_i|\rho_{AB}|x_i\rangle =\mathrm{Tr}_A \left[\rho_{AB}|x_i\rangle \langle x_i| \right] = p_i(x) \left(\frac{1}{N} \mathds{1} + \frac{1}{2} \sum_{\mu=1}^{N^2-1} \vec{r}_{\mu}^{\,(i)}\sigma_{\mu} \right)\; ,
\end{align}
where $p_i(x) = \mathrm{Tr}_B[\rho_{i|x}]$ and $\sigma_{\mu}$ are the generators of SU($N$) group. The measurement $X'$ on $B$'s side gives
\begin{align}
P_{ij}(X,X') & = \langle x'_j| \rho_{i|x}|x'_j\rangle = \mathrm{Tr}_B[\rho_{i|x} |x'_j\rangle \langle x'_j|] \nonumber \\
& = p_i(x) \left( \frac{1}{N} + \frac{1}{2} \vec{r}^{\,(i)}\cdot \vec{t}^{\,(j)} \right) \; .
\end{align}
Here $|x_j'\rangle\langle x_j'|=\frac{1}{N}\mathds{1} + \frac{1}{2} \vec{t}^{\,(j)}\cdot \vec{\sigma}$ is the Bloch representation of the projection measurement for $B$. In this sense, the maximization of the elements of $P(X,X')$ turns to linear maximization over the Bloch vector $\{ \vec{t}^{\,(j)} \}$. The Bloch vectors for quantum states have specific constrains \cite{Positivity-Bloch}. For the the complete basis of $|x_j'\rangle \langle x_j'|$, we further have $\sum_{j=1}^N \vec{t}^{\,(j)} =0$. We define $N$ nonempty sets of natural numbers, i.e., $I_j^{(k)} \subseteq \{1,\cdots, N\}$ where $1\leq j,k\leq N$ and $k$ stands for the cardinality of the sets. Let the maximal sum of the first $k$ components for $\vec{p}(x|x')$ in Corollary 1 be $\wp_k$, we have
\begin{align}
\wp_k = \max_{\substack{\{ \vec{t}^{\,(1)},\cdots,\vec{t}^{\,(N)}\} \\ \{I_{1}^{(k)},\cdots, I_{N}^{(k)}\}}}\left\{ \sum_{j=1}^{N}\left[ \sum_{i\in I_{j}^{(k)}} \left( \frac{ p_{i}}{N}+\frac{p_{i}}{2}\vec{r}^{\,(i)}\cdot \vec{t}^{\,(j)}\right) \right] \right\} \; . \label{S-Maximal-vector}
\end{align}
For each number $k$, there will be a set of measurement bases $\{\vec{t}^{\,(1)}, \cdots, \vec{t}^{\,(n)}\}$ (in Bloch vector form) that gives the maximal value for equation (\ref{S-Maximal-vector}). This provides a measurement for $B$ which constitutes the optimal measurement strategy $\{X'^{(k)}\}$.

We present a calculation to show how equation (\ref{S-Maximal-vector}) works for the following quantum states and observables
\begin{align}
|\psi\rangle & = \cos\xi |00\rangle + \sin\xi |11\rangle \; , \\
\sigma(\theta,\phi) & =\begin{pmatrix}
\cos\theta & e^{-i\phi} \sin\theta \\
e^{i\phi}\sin\theta & -\cos\theta
\end{pmatrix} \; ,
\end{align}
with $\xi\in[0,\pi/4]$ and $\theta \in [0,\pi]$. We choose the measurement $X=\sigma(\theta,\phi)$, and the resulted assemblages for $B$ are
\begin{align}
\rho_{1|x} & = \frac{1+\cos\theta\cos(2\xi)}{4} \mathds{1} + \frac{\sin\theta\sin(2\xi)\cos\phi}{4} \sigma_x + \frac{\sin\theta\sin(2\xi)\sin\phi}{4} \sigma_y + \frac{\cos\theta+\cos(2\xi)}{4}\sigma_z \; ,\\
\rho_{2|x} & = \frac{1-\cos\theta\cos(2\xi)}{4}\mathds{1} - \frac{\sin\theta\sin(2\xi)\cos\phi}{4}\sigma_x - \frac{\sin\theta\sin(2\xi)\sin\phi}{4}\sigma_y - \frac{\cos\theta -\cos(2\xi)}{4}\sigma_z \;.
\end{align}
Here $p_{1}(x) = \frac{1+\cos\theta\cos(2\xi)}{2}$ and $p_{2}(x) = \frac{1-\cos\theta\cos(2\xi)}{2}$. The qubit Bloch vectors for $|x_j'\rangle\langle x_j'|$ take the form of
\begin{align}
|x_1'\rangle\langle x_1'| = \frac{1}{2}\mathds{1} + \frac{\vec{t}\cdot\vec{\sigma}}{2} \; , \;|x_2'\rangle\langle x_2'| = \frac{1}{2}\mathds{1} - \frac{\vec{t}\cdot\vec{\sigma}}{2} \; ,
\end{align}
where $\vec{t}^{\,(1)} = \vec{t} =-\vec{t}^{\,(2)}$. For qubit system we need only to consider the case of $k=1$ in equation (\ref{S-Maximal-vector}) (there are only two elements in $\vec{p}(x|x')$ and $p_2(x|x') = 1-p_1(x|x')$)
\begin{align}
\left\{\begin{array}{l l}
\left\{I_{1}^{(1)}=\{1\},I_{2}^{(1)} =\{1\} \right\} : & \displaystyle \max_{\{\vec{t}\,\}} \left\{ \sum_{j=1}^2 \left[\sum_{i\in I_{j}^{(1)}} \left(\frac{p_{i}}{2} + \frac{p_{i}}{2} \vec{r}^{\,(i)} \cdot \vec{t}^{\,(j)}\right)\right] \right\}\\ \\
\left\{I_{1}^{(1)}=\{1\},I_{2}^{(1)} = \{2\} \right\}: & \displaystyle \max_{\{\vec{t}\,\}} \left\{ \sum_{j=1}^2 \left[\sum_{i\in I_{j}^{(1)}} \left(\frac{p_{i}}{2} + \frac{p_{i}}{2} \vec{r}^{\,(i)} \cdot \vec{t}^{\,(j)}\right)\right] \right\}
\end{array} \right. \;.
\end{align}
Considering $\vec{t}=\vec{t}^{\,(1)}=-\vec{t}^{\,(2)}=(\sin\theta_t\cos\phi_t, \sin\theta_t\sin\phi_t, \cos\theta_t)$, the maximization goes as
\begin{align}
\left\{I_{1}^{(1)}=\{1\},I_{2}^{(1)} =\{1\} \right\}: & \hspace{0.3cm} \max_{\{\vec{t}\,\}} \left\{ \frac{1}{2} + \frac{\cos\theta\cos(2\xi)}{2} \right\}\; , \label{S-qubit-11}\\
\left\{I_{1}^{(1)}=\{1\},I_{2}^{(1)} =\{2\} \right\}: & \hspace{0.3cm} \max_{\{\vec{t}\,\}} \left\{\frac{1}{2}+ \frac{\cos\theta_t\cos\theta+\sin\theta_t\sin\theta\sin(2\xi)\cos(\phi-\phi_t)}{2} \right\}\; . \label{S-qubit-12}
\end{align}
Maximizing over equations (\ref{S-qubit-11}) and (\ref{S-qubit-12}), we have that the maximal value comes from equation (\ref{S-qubit-12}) for $\frac{\cos\theta_t}{\sin\theta_t} = \frac{\cos\theta}{\sin\theta\sin(2\xi)}$ and $\phi_t=\phi$, which gives
\begin{align}
p_1(x|x'^{(1)}) = \frac{1}{2}+ \frac{1}{2}\sqrt{\cos^2\theta + \sin^2\theta\sin(2\xi)} \;.
\end{align}
We finally get
\begin{align}
\vec{s}^{\,(x)} =
\begin{pmatrix}
\displaystyle \frac{1}{2}\left(1+\sqrt{\cos^2\theta + \sin^2\theta\sin^2(2\xi)}\right) \\ \\
\displaystyle \frac{1}{2}\left(1-\sqrt{\cos^2\theta + \sin^2\theta\sin^2(2\xi)}\right)
\end{pmatrix} \; .
\end{align}
Similar result will be obtained for $Y=\sigma(\theta,\pi)$, where
\begin{align}
\vec{s}^{\,(y)} =
\begin{pmatrix}
\displaystyle \frac{1}{2}\left(1+\sqrt{\cos^2\theta + \sin^2\theta\sin^2(2\xi)}\right) \\ \\
\displaystyle \frac{1}{2}\left(1-\sqrt{\cos^2\theta + \sin^2\theta\sin^2(2\xi)}\right)
\end{pmatrix} \; .
\end{align}
And the conditional majorization uncertianty relation becomes
\begin{align}
\vec{p}(x|x') \otimes\vec{p}(y|y') \prec \vec{s}^{\,(\oplus)} = \vec{s}^{\,(x)} \oplus \vec{s}^{\,(y)}
\end{align}
The majorization uncertainty relation for single particle state is \cite{S-Maj-Un-Relation}
\begin{align}
\vec{p}(x) \oplus \vec{p}(y) & \prec \vec{s} =
\begin{pmatrix}
1 \\
\cos\theta \\
2\sin^2\frac{\theta}{2} \\
0
\end{pmatrix} \; . \label{S-maj-pure}
\end{align}
The upper bound of this uncertainty relation $\vec{s}$ has been compared with the $\vec{s}^{\,(\oplus)}$ of conditional majorization uncertainty reation in the Figure 1.

\section{Demonstration of Corollary 1}

$\vec{p}(x|x')$ and $\vec{s}^{\,(x)}$ both are composed of nonnegative real numbers and $\vec{p} (x|x')\prec \vec{s}^{\,(x)}$. The converse of the Schur's Theorem ( equation (II.14) of Ref. \cite{S-Book-Bhatia}) states that there always exist a positive semi-definite Hermitian matrix $\mathcal{X}$ such that
\begin{align}
\vec{p} (x|x')= \{\mathcal{X}_{11},\cdots,\mathcal{X}_{NN}\} \; , \; \vec{s}^{\,(x)} = \{\lambda_1(\mathcal{X}),\cdots, \lambda_{N}(\mathcal{X})\}\; .
\end{align}
Here $\mathcal{X}_{ii}$ and $\lambda_{j}(\mathcal{X})$ are the diagonal elements and  eigenvalues of the Hermitian matrix respectively, and we assume that $\lambda_{1}(\mathcal{X})\geq \lambda_{2}(\mathcal{X})\geq \cdots \geq \lambda_{N}(\mathcal{X})\geq 0$. The positive semidefinite Hermitian matrix $\mathcal{Y}$ also exits for $\vec{p} (y|y')\prec \vec{s}^{\,(y)}$. It is easy to verify
\begin{align}
\vec{p} (x|x') \oplus \vec{p} (y|y') & = \mathrm{diag}\{\mathcal{X} \oplus \mathcal{Y}\} \; , \\
\vec{p} (x|x') \otimes \vec{p} (y|y') & = \mathrm{diag}\{\mathcal{X} \otimes \mathcal{Y}\} \; .
\end{align}
That is to say, the direct sum and direct product of the vectors $\vec{p} (x|x')$ and $\vec{p} (y|y')$ are just the diagonal elements of the direct sum and direct product of $\mathcal{X}$ and $\mathcal{Y}$. Both $\mathcal{X} \oplus \mathcal{Y}$ and $\mathcal{X} \otimes \mathcal{Y}$ are Hermitian matrices. Then Schur's Theorem tells
\begin{align}
\vec{p}(x|x') \oplus \vec{p}(y|y') \prec \vec{s}^{\,(\oplus)} = \vec{s}^{\,(x)} \oplus \vec{s}^{\,(y)}\; , \\
\vec{p}(x|x') \otimes \vec{p}(y|y') \prec \vec{s}^{\,(\otimes)} = \vec{s}^{\,(x)} \otimes \vec{s}^{\,(y)} \; ,
\end{align}
where $\vec{s}^{\,(\oplus)}$ and $\vec{s}^{\,(\otimes)}$ are just the eigenvalues of $\mathcal{X} \oplus \mathcal{Y}$ and $\mathcal{X} \otimes \mathcal{Y}$.

Now considering the sum of the two Hermitian matrices $\mathcal{X} + \mathcal{Y}$, the Lidskii's Theorem (Theorem III.4.1 in Ref. \cite{S-Book-Bhatia}) tells that there exists a vector $\vec{s}^{\,(x+y)}$ such that
\begin{align}
\vec{s}^{\,(x+y)} \prec \vec{s}^{\,(x)\downarrow} + \vec{s}^{\,(y)\downarrow} \; .
\end{align}
Here $\vec{s}^{\,(x+y)}$ is composed of the eigenvalues of $\mathcal{X} + \mathcal{Y}$. Again using the Schur's Theorem for Hermitian matrix we have
\begin{align}
\vec{p}(x|x') + \vec{p}(y|y') \prec \vec{s}^{\,(x+y)} \prec \overbrace{\vec{s}^{\,(x)\downarrow} + \vec{s}^{\,(y)\downarrow}}^{\vec{s}^{\,(+)}} \; .
\end{align}
Note that the vector $\vec{s}^{\,(x+y)}$ can be further strengthened using the Horn's inequalities \cite{QL-sep,Horn-Conj-Rev}.

The Hadamard product of two matrix is defined to be $[\mathcal{X}\circ \mathcal{Y}]_{ij} \equiv \mathcal{X}_{ij}\mathcal{Y}_{ij}$, and similarly we may define the Hadamard product of two vectors as $[\vec{a}\circ \vec{b}\,]_k = a_kb_k$. According to Theorem 5.5.4 of Ref. \cite{S-Book-Horn-2} we have
\begin{align}
\vec{s}^{\, (x\circ y)} \prec_{w} \vec{s}^{\,(x)\downarrow} \circ \vec{s}^{\,(y)\downarrow} \; , \label{S-hadmard-prod}
\end{align}
where $\vec{s}^{\, (x\circ y)}$ is composed of the singular values of $\mathcal{X}\circ \mathcal{Y}$. The {\it weak majorization} for two vectors whose components in descending order, $\vec{a} \prec_w \vec{b}$, means that $\forall k\in \{1,\cdots, N\}$, $\sum_{i=1}^k a_i\leq \sum_{j=1}^k b_j$ and the equality is not required for $k=N$ (this is the difference comparing with the ordinary majorization relation $\prec$). As $\mathcal{X}$ and $\mathcal{Y}$ are positive semidefinite Hermitian matrices, the Schur product Theorem (Theorem 5.2.1 of Ref.\cite{S-Book-Horn-2}) asserts that $\mathcal{X} \circ \mathcal{Y}$ is also a positive semidefinite Hermitian matrix. Therefore the diagonal elements of $\mathcal{X}\circ \mathcal{Y}$ majorized by its eigenvalues
\begin{align}
\vec{p}(x|x') \circ \vec{p}(y|y') \prec \vec{s}^{\, (x\circ y)}\;. \label{S-hadmard-prod-S}
\end{align}
Combining the equations (\ref{S-hadmard-prod}) and (\ref{S-hadmard-prod-S}), we get
\begin{align}
\vec{p}(x|x') \circ \vec{p}(y|y') \prec_{w} \vec{s}^{\,(x)\downarrow} \circ \vec{s}^{\,(y)\downarrow}\; .
\end{align}
This indicates that the Hadamard product $\circ$ may invoke a new operational family of weak majorization uncertainty relation parallel to that of $\{*\}$ in Corollary 1.

\section{Demonstration of Corollary 2}

An arbitrary qubit quantum state can be written as
\begin{align}
\rho_{AB} = \frac{1}{4}\left( \mathds{1}\otimes \mathds{1} + \vec{a}\cdot\vec{\sigma} \otimes \mathds{1} + \mathds{1}\otimes \vec{b}\cdot \vec{\sigma} + \sum_{\mu,\nu=1}^3\mathcal{T}_{\mu\nu}\sigma_{\mu}\otimes \sigma_{\nu}\right) \; .
\end{align}
Here $\mathcal{T}$ is called the correlation matrix of the state $\rho_{AB}$. The quantum nonlocalities of quantum states remain the same under the local unitary operations. The bipartite state can be transformed into
\begin{align}
\rho'_{AB} = \frac{1}{4}\left( \mathds{1}\otimes \mathds{1} + \vec{a}'\cdot\vec{\sigma} \otimes \mathds{1} + \mathds{1}\otimes \vec{b}'\cdot \vec{\sigma} + \sum_{i=1}^3\tau_{i}\sigma_{i}\otimes \sigma_{i}\right) \; ,
\end{align}
where $\tau_i$ are the singular values of the correlation matrix. For the projective measurement on each side
\begin{align}
\mathrm{Alice} &: |x_{\pm}\rangle\langle x_{\pm}| = \frac{1}{2}(\mathds{1} \pm \vec{r}\cdot\vec{\sigma})\; ,\\
\mathrm{Bob} \hspace{0.2cm} & : |x'_{\pm}\rangle\langle x'_{\pm}| = \frac{1}{2}(\mathds{1} \pm \vec{t}\cdot\vec{\sigma})\;,
\end{align}
the resulted joint distribution is
\begin{align}
P(X,X') & =
\begin{pmatrix}
P(x_+,x'_+) & P(x_+,x'_-) \\
P(x_-,x'_+) & P(x_-,x'_-)
\end{pmatrix} \nonumber \\
& = \frac{1}{4}
\begin{pmatrix}
1+\vec{a}'\cdot\vec{r} + \vec{b}'\cdot\vec{t} + \vec{r}^{\mathrm{T}}\Lambda_{\tau}\vec{t} &
1+\vec{a}'\cdot\vec{r} - \vec{b}'\cdot\vec{t} - \vec{r}^{\mathrm{T}}\Lambda_{\tau}\vec{t} \\
1-\vec{a}'\cdot\vec{r} + \vec{b}'\cdot\vec{t} - \vec{r}^{\mathrm{T}}\Lambda_{\tau}\vec{t} &
1-\vec{a}'\cdot\vec{r} - \vec{b}'\cdot\vec{t} + \vec{r}^{\mathrm{T}}\Lambda_{\tau}\vec{t}
\end{pmatrix} \; . \label{S-rhoAB-general}
\end{align}
Here $\Lambda_{\tau} = \mathrm{diag}\{\tau_1,\tau_2,\tau_3\}$ and we assume $\tau_1\geq \tau_2\geq \tau_3\geq 0$. From equation (\ref{S-rhoAB-general}), we may conclude that
\begin{align}
\text{Reduce $X$ by $X'$}& : s^{(x)}_1 = \max\left\{\frac{1}{2}(1+\vec{a}'\cdot\vec{r}\,), \max_{\{\vec{t}\,\}} \{\frac{1}{2}(1+\vec{r}^{\mathrm{T}}\Lambda_{\tau}\vec{t}\,)\}\right\} \; , \label{S-s1x} \\
\text{Reduce $X'$ by $X$}& : s^{(x')}_1= \max\left\{\frac{1}{2}(1+\vec{b}'\cdot\vec{t}\,), \max_{\{\vec{r}\, \}}\{\frac{1}{2}(1+\vec{r}^{\mathrm{T}}\Lambda_{\tau}\vec{t}\,)\}\right\} \; . \label{S-s1xp}
\end{align}
Equation (\ref{S-s1x}) corresponds to the conditional distribution $\vec{p}(x|x')$ and equation (\ref{S-s1xp}) corresponds to $\vec{p}(x'|x)$. The maximization over the Bloch vectors are
\begin{align}
\max_{\{\vec{t}\,\}} \left\{ \frac{1}{2}(1+\vec{r}^{\mathrm{T}}\Lambda_{\tau}\vec{t}\,) \right\} & = \frac{1}{2} + \frac{1}{2}\sqrt{\tau_3^2\sin^2\theta_r\cos^2\phi_r+ \tau_2^2\sin^2\theta_r\sin^2\phi_r + \tau_1^2\cos^2\theta_r} \; , \label{S-mixd-BA}\\
\max_{\{\vec{r}\,\}} \left\{\frac{1}{2}(1+\vec{r}^{\mathrm{T}}\Lambda_{\tau}\vec{t}\,) \right\} & = \frac{1}{2} + \frac{1}{2}\sqrt{\tau_3^2 \sin^2\theta_t\cos^2\phi_t+ \tau_2^2\sin^2\theta_t\sin^2\phi_t + \tau_1^2\cos^2\theta_t} \; .
\end{align}
where $\theta_{t,r},\phi_{t,r}$ are the angles of $\vec{t}$ and $\vec{r}$ respectively.

Suppose we perform an infinite number of projective measurement $|x_+\rangle\langle x_+|$ whose Bloch vectors range as $\theta_r\in [0,\pi/2]$ and $\phi_r\in [0,2\pi]$, the quantum mechanical prediction for the maximal value of the sum of the probabilities of getting $|x_{+}(\theta_r,\phi_r)\rangle$ is  \cite{S-Quantum-Magic}
\begin{align}
\max_{|\psi\rangle} \left\{\sum_{\theta_r,\phi_r} p(x_+(\theta_r,\phi_r)) \right\}= \int_{0}^{2\pi} \int_{0}^{\frac{\pi}{2}} \cos^2\frac{\theta}{2}\,\mathrm{d}\Omega = \frac{3\pi}{2} = \varepsilon_1\; .
\end{align}
Here $\varepsilon_1$ is first element of $\vec{\varepsilon}$ in equation (17) for the infinite measurements $|x_+\rangle\langle x_+|$. From equation (\ref{S-mixd-BA}), we have the same value in the presence of quantum memory
\begin{align}
& \int_{0}^{2\pi} \int_{0}^{\frac{\pi}{2}} \frac{1+\sqrt{\tau_3^2\sin^2\theta_r\cos^2\phi_r+ \tau_2^2\sin^2\theta_r\sin^2\phi_r + \tau_1^2\cos^2\theta_r}}{2} \,\mathrm{d}\Omega \nonumber \\
= & \pi + \frac{1}{2} \int_0^{2\pi}\int_0^{\frac{\pi}{2}} \sqrt{\tau_3^2\sin^2\theta_r\cos^2\phi_r+ \tau_2^2\sin^2\theta_r\sin^2\phi_r + \tau_1^2\cos^2\theta_r} \,\mathrm{d}\Omega \nonumber \\
= & \pi + \pi R_{G}(\tau_3^2,\tau_2^2,\tau_1^2) \; \label{S-integrand-one}
\end{align}
with
\begin{align}
R_{G}(\tau_3^2,\tau_2^2,\tau_1^2) \equiv \frac{1}{4\pi}\int_0^{2\pi}\int_0^{\pi} \sqrt{\tau_3^2\sin^2\theta_r\cos^2\phi_r+ \tau_2^2\sin^2\theta_r\sin^2\phi_r + \tau_1^2\cos^2\theta_r} \,\mathrm{d}\Omega \nonumber
\end{align}
which is the Elliptic integral of No. 19.16.3 see \cite{S-Special-Fun-Ell}. Therefore we arrive that if the state $\rho_{AB}$ is non-steerable then the conditional majorization uncertainty relation should not violate the quantum limit, i.e. the following inequality shall be satisfied
\begin{align}
&  \pi + \pi R_{G}(\tau_3^2,\tau_2^2,\tau_1^2) \leq \varepsilon_1 = \frac{3\pi}{2} \nonumber \\
\Rightarrow & R_{G}(\tau_3^2,\tau_2^2,\tau_1^2)\leq \frac{1}{2}\; . \label{S-steering-crit}
\end{align}
The violation of equation (\ref{S-steering-crit}) thus serves as a sufficient condition for quantum steering. The condition could be further strengthened if the integrand in equation (\ref{S-integrand-one}) is replaced with $\vec{s}^{\,(x)}$ in equation (\ref{S-s1x}), because here we neglected the contribution from $\frac{1}{2}(1+\vec{a}'\cdot\vec{r}\,)$ in equation (\ref{S-s1x}).

\section{The example of bipartite state $\rho_{\xi}$}

The bipartite state $\rho_{\xi} =\displaystyle \frac{1-p}{2}\rho^{A}_{\xi} \otimes \mathds{1} + p|\psi_{\xi}\rangle \langle \psi_{\xi}|$, written in the Bloch vector form, is
\begin{align}
\rho_{\xi} & = \frac{1}{4}(\mathds{1}\otimes \mathds{1} + \cos(2\xi)\sigma_z\otimes\mathds{1}+ \mathds{1} \otimes p\cos(2\xi)\sigma_z + \nonumber \\
& \hspace{1cm} p\sin(2\xi)\sigma_x \otimes \sigma_x -p\sin(2\xi)\sigma_y \otimes \sigma_y+ p\sigma_z \otimes \sigma_z)\;.
\end{align}
According to equation (\ref{S-rhoAB-general}), the joint distribution matrix would be
\begin{align}
& P(X,X') = \nonumber \\ &\frac{1}{4}
\begin{pmatrix}
1+ \cos(2\xi)\cos\theta_r + p\cos(2\xi)\cos\theta_t + \tau_{rt} &
1+\cos(2\xi)\cos\theta_r - p\cos(2\xi)\cos\theta_t- \tau_{rt} \\
1-\cos(2\xi)\cos\theta_r + p\cos(2\xi)\cos\theta_t - \tau_{rt} &
1-\cos(2\xi)\cos\theta_r - p\cos(2\xi)\cos\theta_t + \tau_{rt}
\end{pmatrix}\;.
\end{align}
Here
\begin{align}
\tau_{rt} & = p\left[\sin(2\xi) \left( \sin\theta_r \cos\phi_r\sin\theta_t\cos\phi_t -\sin\theta_r \sin\phi_r\sin\theta_t\sin\phi_t \right) +\cos\theta_r\cos\theta_t \right]\nonumber \\
& =  p\left[\sin(2\xi) \sin\theta_r \sin\theta_t\cos(\phi_r-\phi_t) +\cos\theta_r\cos\theta_t \right] \; . \label{S-tau-rt}
\end{align}
For measurement $X$ the optimal measurement $X'$ is given by maximizing the following
\begin{align}
s_1^{\,(x)} = \max\left\{ \frac{1+\cos(2\xi) \cos\theta_r}{2},\max_{\vec{t}} \{ \frac{1+\tau_{rt}}{2} \} \right\} \;, \label{S-s-B-steer-A}
\end{align}
while for measurement $X'$ the optimal measurement $X$ is given by
\begin{align}
s_1^{\,(x')} = \max\left\{ \frac{1+p\cos(2\xi)\cos\theta_t}{2},\max_{\vec{r}} \{ \frac{1+\tau_{rt}}{2} \} \right\}\;. \label{S-s-A-steer-B}
\end{align}
Here it is easy to see from equation (\ref{S-tau-rt}) that
\begin{align}
\max_{\vec{t}}\{\frac{1+\tau_{rt}}{2}\} & = \frac{1+p\sqrt{\cos^2\theta_r+\sin^2\theta_r\sin^2(2\xi)}}{2} \; , \text{where} \, \tan\theta_t = \tan\theta_r\sin(2\xi) \; ,  \\
\max_{\vec{r}}\{\frac{1+\tau_{rt}}{2}\} & = \frac{1+p\sqrt{\cos^2\theta_t+\sin^2\theta_t\sin^2(2\xi)}}{2} \; , \text{where} \, \tan\theta_r = \tan\theta_t\sin(2\xi) \; ,
\end{align}
where $\theta_r=\theta_t$ for both equations. Now equations (\ref{S-s-B-steer-A}) and (\ref{S-s-A-steer-B}) turn to
\begin{align}
s_1^{\,(x)} & = \max\left\{ \frac{1+\cos(2\xi) \cos\theta_r}{2},  \frac{1+p\sqrt{\cos^2\theta_r+\sin^2\theta_r\sin^2(2\xi)}}{2} \right\} \;, \label{S-B-steer-A} \\
s_1^{\,(x')} & = \max\left\{ \frac{1+p\cos(2\xi)\cos\theta_t}{2}, \frac{1+p\sqrt{\cos^2\theta_t+\sin^2\theta_t\sin^2(2\xi)}}{2} \right\}\;. \label{S-A-steer-B}
\end{align}
Equation (\ref{S-A-steer-B}) gives $s_1^{(x')} = \frac{1+p\sqrt{\cos^2\theta_t+\sin^2\theta_t\sin^2(2\xi)}}{2}$, and the optimal measurement is
\begin{align}
\text{For}\; \cos\theta_t\in[0,1]\;, \; \tan\theta_r = \tan\theta_t\sin(2\xi)\; \text{and} \; \theta_r=\theta_t \; .
\end{align}
This is just the equation (22) in the main text, and equation (21) can be similarly obtained from equation (\ref{S-B-steer-A}).

\end{document}